# Realizing highly entangled states in asymmetrically coupled three NV centers at room temperature

*Declan Mahony and Somnath Bhattacharyya[a)]*

*Nano-Scale Transport Physics Laboratory, School of Physics, University of the Witwatersrand, Private Bag 3, WITS 2050, Johannesburg, South Africa*

Despite numerous efforts the coupling between randomly arranged multi-NV centers and also resonators has not been improved significantly mainly due to our limited knowledge of their entanglement times ($2\tau_{ent}$). Here, we demonstrate a very strong coupling between three-NV centers by using a simulated triple electron-electron resonance experiment based on a new quantum ($U_C$) gate on IBM quantum simulator with $2\tau_{ent}$ ~12.5 $\mu s$ arranged is a triangular configuration. Interestingly through breaking the symmetry of couplings an even lower $2\tau_{ent}$ ~6.3 $\mu s$ can be achieved. This simulation not only explains the luminescence spectra in recently observed three-NV centers [Haruyama, Nat. Commun. 2019] but also shows a large improvement of the entanglement in artificially created structures through a cyclic redistribution of couplings. Realistically disordered coupling configurations of NV centers qubits with short time periods and high (0.89-0.99) fidelity of states clearly demonstrate possibility of accurate quantum registers operated at room temperature.

The concept of simulating quantum mechanical systems more efficiently on quantum computers than on classical computers has become far more realisable in recent years with the development of quantum registers consisting of up to tens of superconducting qubits [1]. This has already enabled the effective simulations of the dynamics of many-body quantum mechanical systems [2, 3], condensed matter physics [4, 5, 6], high-energy physics [7, 8] and quantum chemistry [9, 10]. Simulating complex quantum mechanical systems requires complex quantum circuits consisting of more quantum gates. Such circuits take a longer time to operate on qubits, so ideally the coherence time of the quantum register should be much longer than the operating time of the quantum circuit since the accuracy of quantum simulation depends on the coherence time of the qubits [11]. Currently, a single flux qubit can have coherence times up to 0.5 ms [12], while a spin qubit consisting of a divacancy in silicon carbide can reach 1.3 ms [13]. A silicon-vacancy spin qubit can demonstrate coherence times of up to 13 ms and spin relaxation times up to 1 s below 500 mK [14], however a quantum computer that requires such low temperatures is a very limiting factor - ideally the most practical quantum computer should be able to operate at room temperature [15]. NV centers in diamond show the most promise in this regard since they have demonstrated the longest coherence times at room temperature compared to other defects in diamond with a natural abundance (1.1%) of $^{13}C$ - 0.7 ms. This coherence time can be improved to 1.8 – 2.0 ms by suppressing impurities and defects [16]. This makes NV centers a good candidate for spin qubits in a quantum computer (theoretically) at room temperature that could provide a distinct advantage over superconducting flux qubits [17]. However, in practice, a three NV center quantum register has not been a feasibly achievable option due to the difficulty of fabricating three coupled NV centers in diamond. Recently, Haruyama *et al*. have claimed the synthesis and analysis of three coupled NV centers using implantation of $C_5N_4H_n$ ions from an adenine source to scale up the creation of NV centers in diamond [18]. Not all NV centers that were close enough (in the order of 10 nm) were coupled though only one group of three NV centers was classified as strongly coupled, and this triplet group was further discussed which needs further theoretical support [18].

The entanglement of quantum states is essential to the formalism of quantum theory [19], and the entanglement of qubits in a quantum register is fundamental to the operation of quantum circuits that can theoretically outperform classical computers when it comes to quantum simulation [20]. Entanglement of two NV centers at room temperature has been demonstrated [17], but until recently the large-scale creation of three coupled NV centers in diamond has proven difficult [18]. Here, we show the performance of newly developed quantum gates that enables quantum simulation of the entanglement of first two, and then three such coupled NV centers using a double and triple electron resonance pulse sequence [21, 22]. This operation not only explain the experimental data in Ref. 18 but also demonstrate a very short entanglement time in an ordered configuration.

A triangular configuration of NV centers is important for creating an extended lattice structure as used in superconducting qubits. This will enable simulations of many body interactions in the presence of disorder or unequal coupling between qubits or spin centers [2-6]. Therefore, we extend the work beyond ordered configurations of NV centers by breaking the symmetry to produce strong luminescence (resonance) features. The coupling strength between any two NV center defects depends on the size of the quantum dots as well as the distance between them [23]. Like the resonance spectra of quantum dots the geometric configurations of the NV centers in relation to one another (such as vertical, horizontal or triangular) would affect the luminescence if the coupling between them are varied [24]. Simulating the entanglement of an idealised configuration of three coupled NV centers has value, but the demonstration of entanglement in a non-ideal system is more relevant with respect to the physical realisation of a three NV center quantum register, as physically realised systems are rarely ideal. Considering this, the distance between NV centers and the geometry of the configuration is taken to resemble that of a triangle in terms of the coupling between the NV centers, as these configurations are the most disordered. The parameter under investigation here is simply coupling strength for different ordered and disordered configurations of the NV centers. The asymmetry of a given configuration is represented by the use of a coupling constant (R) defined by

$$R = \left(\frac{1}{v_{AC}} + \frac{1}{v_{AB}} + \frac{1}{v_{CB}}\right)^{-1} \quad (1)$$

where $v_{AC}, v_{AB}$ and $v_{CB}$ are the coupling strengths between the three NV centers as shown in Figure 1. In particular, three extreme configurations are examined – equilateral representation with equal coupling strengths between each NV center ($v_{AB} = v_{AC} = v_{CB}$), isosceles representation with equal coupling strengths between two of the three NV centers ($v_{ik} = v_{ij} \neq v_{jk}$) for $i, j, k \; \epsilon \{A, B, C\}$, and scalene representation with different coupling strengths between each NV center ($v_{AB} \neq v_{AC} \neq v_{CB}$). Since the

[a)] Electronic mail: somnath.bhattacharyya@wits.co.za

triple electron resonance scheme developed here to entangle three coupled NV centers requires the decoupling of one pair of NV centers in the final free evolution period, the rotation of each of these configurations is also investigated. This is achieved by a cyclic redistribution of the coupling strengths of each coupled NV center pair. To demonstrate the physical importance of these simulations, the scalene configuration is investigated by using the coupling strengths found by Haruyama et al. [18].

A single negatively charged NV center defect (Figure 1.a.) consists of six electrons existing in a spin-triplet state with the energy level splitting $m_s = 0$ and $m_s = \pm 1$ controlled by external magnetic fields, shown in Figure 1.b. This allows the useful formation of a two-level $m_s = 0$ and $m_s = 1$ system that can act as a single qubit. The NV center defect interacts with the nuclei of surrounding nitrogen and $^{13}C$ atoms in the diamond. This is a major source of noise that creates decoherence in NV centers [25], but for $^{12}C$ enriched diamond, electron spin coherence time can reach 3.0 ms [17]. Even with the more conservative coherence time of 2.0 ms, NV center electron spin qubits have long enough coherence times for accurate execution of fairly complex quantum circuits. The double and triple electron resonance pulse schemes include a free evolution of the system of NV centers for a time of the order of $10^2$ μs, so the noise from electron-nuclear spin interactions is negligible. This means that the electron-nuclear hyperfine coupling terms are excluded from the secular Hamiltonian describing the coupling between three NV center electron spins. The interaction between electron spins of different NV centers can be simulated in a similar way as the interaction between nuclear and electron spin in one NV centers [26]. In a coupled NV center triplet (comprised of $NV_A, NV_B$ and $NV_C$), each NV center is coupled to the other two NV centers, as shown in Figure 1.c).

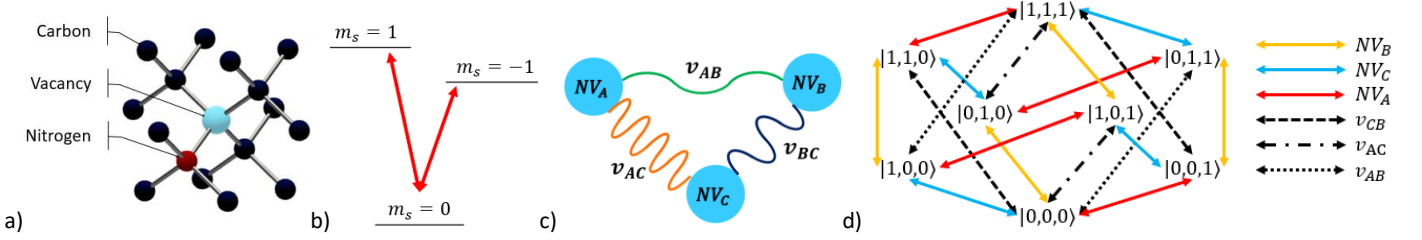

Figure 1. a) Visual representation of a negative NV center defect in diamond. b) Energy level transition of the NV center electron spin that can be controlled by an external magnetic field. c) Three strongly coupled NV centers. $NV_A$ and $NV_C$ are the most strongly coupled, while $NV_A$ and $NV_B$ are the least strongly coupled [18]. d) Effective combined energy level scheme of the three coupled NV centers, where each NV center forms the two-level $m_s = 0$ and $m_s = 1$ system. Spin transitions represented by solid lines can be driven with microwaves due to the different Zeeman shift caused by different magnetic field alignments for each NV center [17]. Transitions represented by dashed lines are driven by the dipolar coupling between NV centers.

Using the secular approximation, the system in a magnetic field B is described by

$$H = \sum_{i=A}^{C}[\Delta S_i^2 + \gamma_e \hat{B}. \hat{S}_i] + H_{dip} \quad (2)$$

where $\Delta = 2.87\ GHz$ is the zero-field splitting [27], $\gamma_e = 2.8\ MHz/G$ is the gyromagnetic ratio [28] and $\hat{S}_i$ is the spin operator for i∈{A,B,C}. The last term $H_{dip}$ is the dipolar coupling term between the three NV centers, given by:

$$H_{dip} = v_{AB}\ \hat{S}_{zA}\ \hat{S}_{zB} + v_{AC}\ \hat{S}_{zA}\ \hat{S}_{zC} + v_{CB}\ \hat{S}_{zC}\ \hat{S}_{zB} \quad (3)$$

The dipolar coupling between NV center spins is $v_{AC}$ between centers A and C, $v_{CB}$ between centers C and B and $v_{AB}$ between centers A and B. The spin-flip terms $\hat{S}_{xi}\hat{S}_{xj} + \hat{S}_{yi}\hat{S}_{yj}$ for i ∈ {A,B,C}, j ∈ {B,C,A} are ignored because the dipolar coupling is smaller than the energetic detuning between any two spins [17]. $H_{dip}$ is considered when simulating the coupling between NV centers in the absence of an external magnetic field, with spin-flip transitions of individual NV centers driven by microwave π pulses as shown in Figure 1.d.

In general, the circuit representing the free evolution of two coupled NV centers (i and j) is found by representing the two-qubit gate $U_C = e^{i\ 2\tau\ v_{ij}(Szi \otimes Szj)}$ with an $Rz(2\tau\ v_{ij})$ gate on qubit $q_i$ and a $C_{q_i}NOT_{q_j}$ gate, with Hadamard gates initially applied to each qubit to allow for interaction. $2\tau$ is the free evolution time – the time period over which the NV centers evolve freely, and $v_{ij}$ is the dipolar coupling between the NV centers. Using the $U_C$ gate, a quantum circuit is developed to simulate a DEER experiment for each coupled pair in a system of three coupled NV centers - $NV_A$, $NV_B$ and $NV_C$ with dipolar coupling $v_{AB} = 4.6\ kH, v_{AC} = 53\ kH$ and $v_{CB} = 24.1\ kH$. The results of the simulated DEER experiment are shown in Figure 2. Here, the simulated luminescence of each NV center for a given $2\tau$ in a DEER experiment is shown on the vertical axis by the normalised counts: the proportion of measurements that resulted in the labelled state out of a total of 1024 measurements for each $2\tau$. The normalised counts are based off the ground state, as the excited state changes depending on which sensor-emitter pair is being simulated. Figure 2 shows the oscillation between the ground state and the excited state that would be physically observed in a real DEER experiment [17, 18]. The dipolar coupling between any two NV centers is a direct measurement of the period of oscillation of each of the plotted lines in Figure 2, where normalised counts are plotted against $\tau$ instead of $2\tau$ to compare with the resuts found by Haruyama et al. The DEER plots are simulated reproductions of the intensity plots that Haruyama et al. used to show the fabrication of three strongly coupled NV centers [18].

By modifying the DEER quantum circuit for two coupled NV centers, a new entanglement circuit is developed (see Methodology). The entanglement of two coupled NV centers with dipolar coupling $v = 4.93\ kHz$ is analysed by measuring the normalised counts of all possible states of the two NV center qubits over increasing free evolution times. The evolution time required for entanglement is defined as $2\tau_{ent}$. From Figure 2, it can be seen that the two qubits reach the entangled Bell state $\frac{1}{\sqrt{2}}(|00\rangle + |11\rangle)$ after a time $2\tau_{ent} = 21.2$ μs for the Quasm Simulator and $2\tau_{ent} = 22.0$ μs for the IBM London quantum emulator. These values are comparable to the $2\tau_{ent} = 25.0$ μs found by Dolde et al. [17]. A triple electron resonance scheme [21, 22] was used to entangle three NV centers, with each NV center acting as the sensor and emitter for the other two NV centers. The linear combination of the states of all three NV centers is considered as the overall state of the system. The decoupling of $NV_B$ and $NV_C$ [30] for the final period of free interaction just before measurement allows the entanglement of the three NV centers. To properly analyse the entanglement, the plots of the relevant states are measured for each case. The coupling between three NV centers results in each NV center qubit oscillating between the ground $|0\rangle$ and excited $|1\rangle$ state according to the linear combination of the oscillation caused by the individual coupling of the NV center with each of the other two NV centers. The proportional occupation (normalised counts) of each state of the system follows similar linear combinations of oscillations caused by the coupling between NV centers. The fidelity of the system over time is calculated by $F = 2\ tr(\rho\sigma)$ where $\rho$ is the density matrix of the target entangled state and $\sigma$ is the density matrix of the measured state for increasing $2\tau$. Fidelity is also plotted over increasing $2\tau$ as a means of further justifying the time at which the system reaches the entangled state, as the maximum fidelity shows when the system becomes closest to the entangled state. The real part of the density matrix of the system at $2\tau_{ent}$ provides a visual representation of how close the system comes to the entangled state. We simulated 27 different configurations of the three coupled NV centers, the 12 important results of which are summarised in Table 1. The most interesting of these configurations are discussed further.

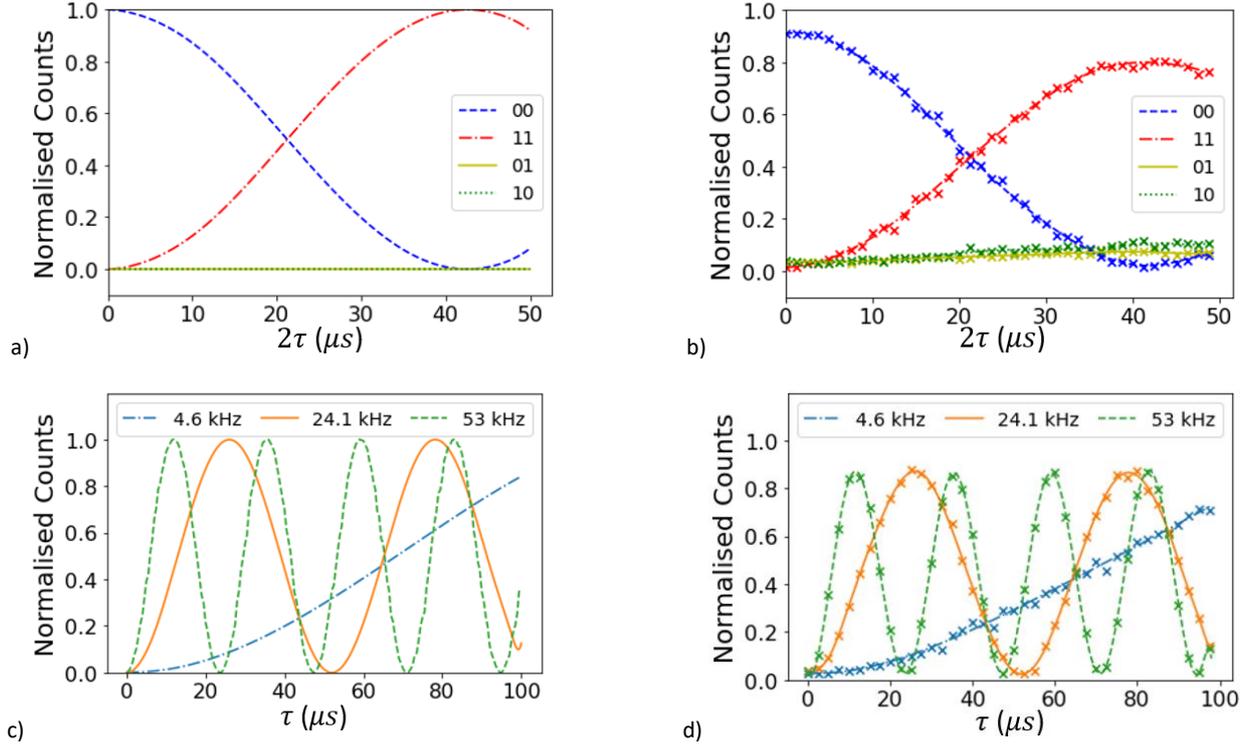

Figure 2: The entanglement of two NV centers with dipolar coupling of $v = 4.93\ kHz$ at a) $2\tau_{ent} = 21.7$ μs using the IBM Quasm Simulator and at b) $2\tau_{ent} = 22.0$ μs using a two qubit register of the IBM London Quantum Emulator [29]. The Normalised Counts represent the relative intensity of light measured. The plotted results of simulated DEER experiments on three coupled NV centers with varying dipolar coupling strengths using a three qubit register of c) the IBM Quasm Simulator and d) the IBM London Quantum Emulator [29]. The Normalised Counts represent the relative intensity of light measured.

The first configuration of three coupled NV centers is the symmetrical case, with an equilateral (type) representation of NV centers where $v_{AB} = v_{AC} = v_{BC}$, inset in Figure 3.b. The entangled state reached for this configuration is $\frac{1}{\sqrt{2}}(|000\rangle - |111\rangle)$ as shown by the density matrix in Figure 3.b. The $|000\rangle$ and $|111\rangle$ states are plotted to show how the system evolves over increasing $2\tau$. Three different coupling strengths are analysed for this equilateral representation. The shape of the evolution of the $|000\rangle$ and $|111\rangle$ states for each of these different coupling strengths is the same (Figure 3.a), suggesting that there could be a consistent evolution of the system for the equilateral representation. Importantly, $2\tau_{ent}$ is shorter for strong coupling and longer for weak coupling.

The second configuration of coupled NV centers is the isosceles (type) representation with twelve different arrangements of coupling strength classified as either single dominant ($v_{ij} = v_{ik} < v_{jk}$) or double dominant ($v_{ij} = v_{ik} > v_{jk}$) with $i, j, k$ iterating through $A, B, C$. Table 1 shows that when $v_{CB} \neq v_{AB} = v_{AC}$, $2\tau_{ent}$ is shorter than other arrangements, with the exception for the case where $v_{CB} \ll v_{AB} = v_{AC}$. As expected, there is a symmetry in different configurations with the same coupling strength for $v_{BC}$: interchanging the coupling strengths of $v_{AB}$ and $v_{AC}$ results in the same $2\tau_{ent}$, albeit with two different states being entangled. Interestingly, configurations with a larger difference between maximum and minimum coupling strengths reached an entangled state after a shorter $2\tau$. Double dominant configurations generally have shorter $2\tau_{ent}$, except for when $v_{CB} \gg v_{AB} = v_{AC}$, which provides an interesting anomaly that is further discussed. The evolution of the $|100\rangle$ and $|011\rangle$ states as well as the fidelity to this entangled state is shown in Figure 3.a. The maximum fidelity of 0.963, as well as the point at which the normalised counts for the $|100\rangle$ state equals the normalised counts of the $|011\rangle$ state, occurs at 6.3 $\mu$s. After a free evolution time of 6.3 $\mu$s, the configuration with $v_{AB} = v_{AC} = \ll v_{CB}$ (Figure 3.a. inset) became very close to the entangled state $\frac{1}{\sqrt{2}}(|100\rangle + |011\rangle)$, shown in Figure 3.d. This shows promise for the entanglement of NV center spin qubits with a more disordered configuration.

The third configuration of coupled NV centers is the scalene (type) representation, with six different arrangements of $v_{ij} \neq v_{ik} \neq v_{jk}$, with $i, j, k$ iterating through $A, B, C$ (Figure 4.b. inset). The $2\tau_{ent}$ values are longer than for the previous two configurations, which is expected as this is the most distorted system that has been simulated here. Each coupled NV center pair was taken to have a different coupling strength than the other two coupling pairs, with coupling strength taking values of 5 $kHz$, 20 $kHz$ or 50 $kHz$. Some interesting similarities in entanglement times were observed, namely that the system reached different entangled states in $2\tau_{ent} = 62.8$ $\mu$s when $v_{CB} = 5$ $kHz$, and in $2\tau_{ent} = 125.7$ $\mu$s otherwise. The evolution of the relevant states for the interesting configurations are shown over increasing $2\tau$ in Figure 4.a and 4.c, with the density matrix of the entangled states reached shown in Figure 4.b and 4.d. From these figures, it can be seen that interchanging the coupling strengths $v_{AC}$ and $v_{AB}$ changes the entangled state reached in the same way as interchanging the first two qubits in the quantum circuit.

A realistic case of the scalene representation uses the measured values of the coupling strengths between three coupled NV centers as found by Haruyama et al. [18]. As before, there are six possible arrangements of this set of coupling strengths, the results of three of which are summarised in Table 1. With all arrangements of these coupling strengths, the entangled state $\frac{1}{\sqrt{2}}(|000\rangle - |111\rangle)$ was reached at $2\tau_{ent} = 130.1$ $\mu$s, however in the cases with $v_{CB} = 24.1$ $kHz$, the entangled Greenberger-Horne-Zeilinger (GHZ) state was reached at $2\tau_{ent} = 11.7$ $\mu$s, as shown in Figure 5. Importantly, the longest and shortest $2\tau_{ent}$ for this disordered configuration is comparable to the longest and shortest $2\tau_{ent}$ for both the weakly coupled equilateral and single dominant isosceles representations. This implies that disordered configurations are useful for the development of an NV center quantum register. Interchanging the coupling $v_{AC}$ and $v_{AB}$ changes the entangled GHZ state from $\frac{1}{\sqrt{2}}(|000\rangle + |111\rangle)$ to $\frac{1}{\sqrt{2}}(|001\rangle + |110\rangle)$.

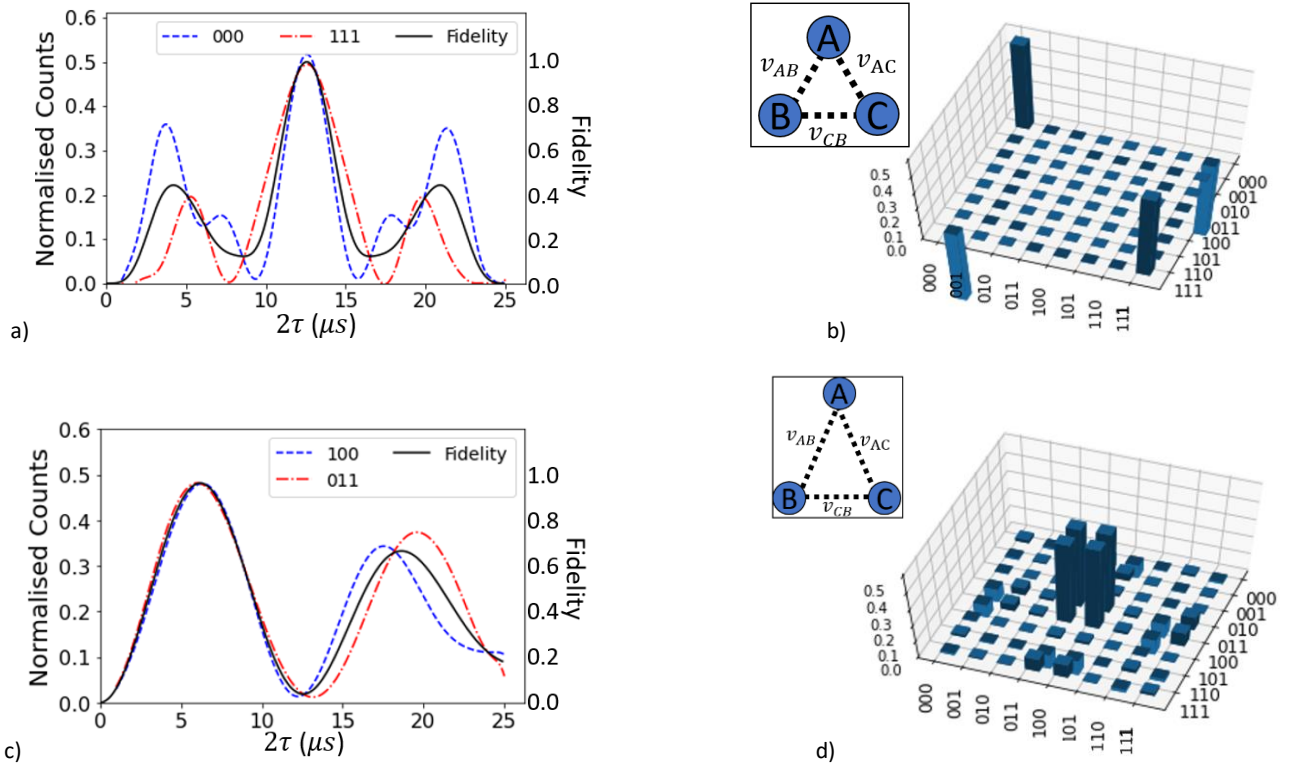

Figure 3: Entanglement of equilateral-type and single dominant isosceles-type representations. a) Evolution of the relevant $|000\rangle$ and $|111\rangle$ states for increasing $2\tau$ for strong coupling of $v_{AB} = v_{AC} = v_{CB} = 50$ $kHz$ with entanglement at $2\tau_{ent} = 12.5$ $\mu$s. b) The entangled GHZ state $\frac{1}{\sqrt{2}}(|000\rangle - |111\rangle)$ reached for the equilateral representation of NV centers (inset) with $v_{AB} = v_{AC} = v_{CB}$. The evolution graphs of the relevant $|000\rangle$ and $|111\rangle$ states for weak and moderate coupling have the same shape stretched over a longer free evolution time. c) Evolution of the relevant $|100\rangle$ and $|011\rangle$ states for increasing $2\tau$ for coupling of $v_{AB} = v_{AC} = 5$ $kHz < v_{CB} = 50$ $kHz$ with entanglement at $2\tau_{ent} = 6.3$ $\mu$s. d) The entangled state $\frac{1}{\sqrt{2}}(|100\rangle + |011\rangle)$ reached for the single dominant isosceles representation of three NV centers (inset) with $v_{AB} = v_{AC} = 5$ $kHz < v_{CB} = 50$ $kHz$.

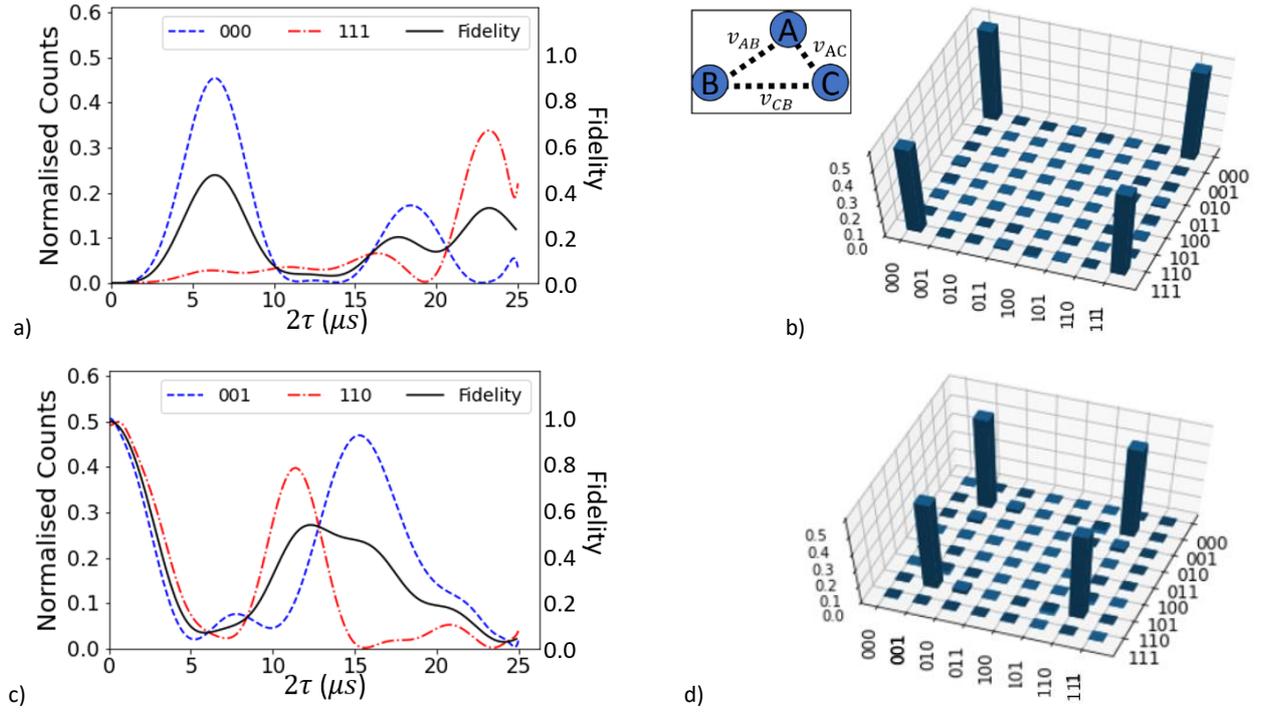

Figure 4: Entanglement of arbitrary scalene-type representation. a) The evolution of the relevant $|000\rangle$ and $|111\rangle$ states for increasing $2\tau$ for scalene coupling configuration of $v_{AC} = 50\ kHz$, $v_{AB} = 20\ kHz$, $v_{CB} = 5\ kHz$ with b) entangled GHZ state $\frac{1}{\sqrt{2}}(|000\rangle + |111\rangle)$ reached at 62.8 $\mu$s. c) The evolution of the relevant $|001\rangle$ and $|110\rangle$ states for increasing $2\tau$ for scalene coupling representation of $v_{AC} = 20\ kHz$, $v_{AB} = 50\ kHz$, $v_{CB} = 5\ kHz$ with d) entangled state $\frac{1}{\sqrt{2}}(|001\rangle + |110\rangle)$ reached at 62.8 $\mu$s. Interchanging the coupling $v_{AC}$ and $v_{AB}$ changes the entangled GHZ state from $\frac{1}{\sqrt{2}}(|000\rangle + |111\rangle)$ to $\frac{1}{\sqrt{2}}(|001\rangle + |110\rangle)$.

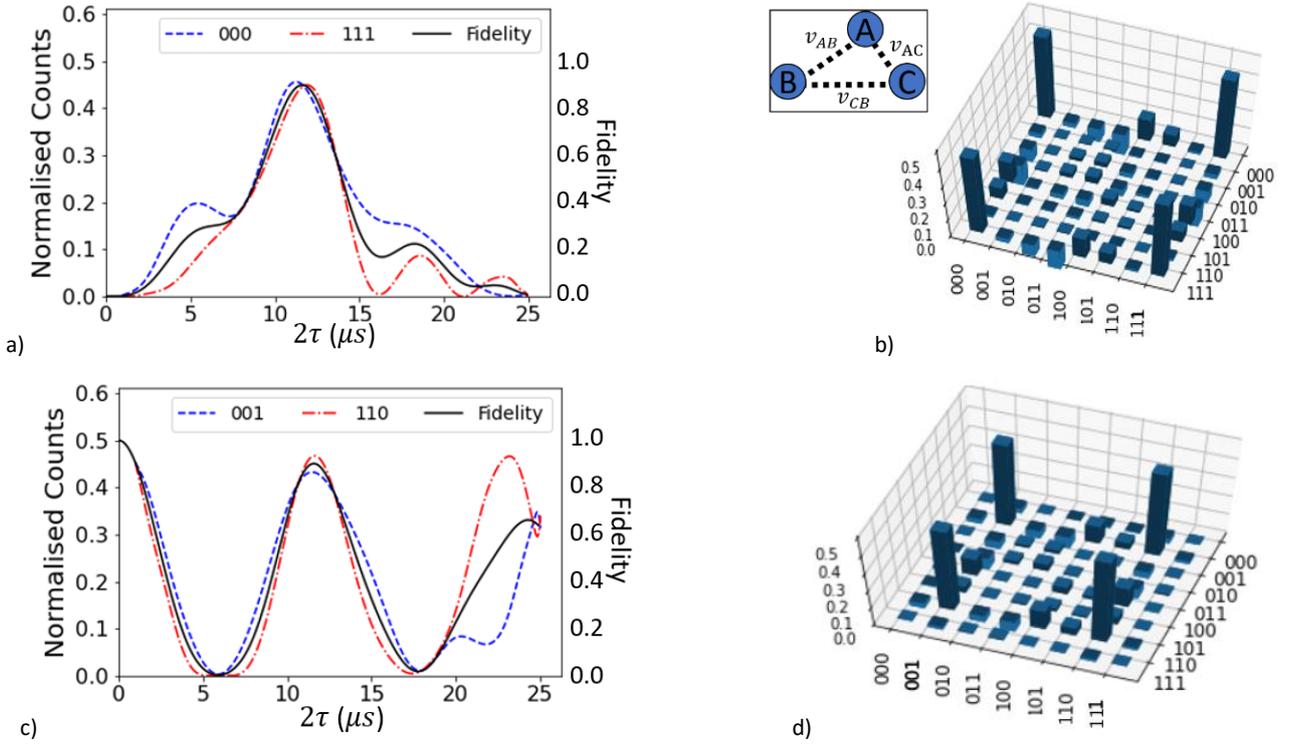

Figure 5: Entanglement of realistic scalene-type configuration. a) The evolution of the relevant $|000\rangle$ and $|111\rangle$ states for increasing $2\tau$ for realistic scalene coupling configuration where $v_{CB} = 24.1\ kHz$ with b) entangled GHZ state $\frac{1}{\sqrt{2}}(|000\rangle + |111\rangle)$ reached at 11.7 $\mu$s. c) The evolution of the relevant $|001\rangle$ and $|110\rangle$ states for increasing $2\tau$ for the same scalene configuration where $v_{AC}$ and $v_{AB}$ are interchanged with d) entangled state $\frac{1}{\sqrt{2}}(|001\rangle + |110\rangle)$ reached at $2\tau_{ent} = 11.7\ \mu$s.

The triple electron resonance circuit (see Methodology) theoretically entangles the three qubits in the $\frac{1}{\sqrt{2}}(|001\rangle - |110\rangle)$ state with $2\tau = 0$, however different entangled states can be achieved with relatively high fidelity for a free evolution time of $2\tau_{ent} = n\pi/v_{max}$ for some integer $n$ where $v_{max}$ is the strongest coupling in the configuration. For equilateral representations, a maximally entangled state was reached for $n = 2$ with a fidelity of 0.996. For the isosceles representation, the evolution of the system is dominated by the more strongly coupled NV centers, so the three qubits can reach an entangled state for $n = 2$ as with the equilateral representation, but with notably lower fidelity. The effect of one pair of NV centers having a different coupling strength results in variations in $2\tau_{ent}$, with the longest case being $n = 10$. The shortest case ($n = 1$) was due to the strongly coupled pair being decoupled for the final free evolution period, thus decreasing decoherence for short $2\tau$. The scalene representation consistently reached an entangled state for $n = 10$ and $n = 11$ with relatively high fidelity ranging from 0.897 to 0.990 for both the idealised and realistic cases respectively, however shorter $2\tau_{ent}$ were observed.

In the idealistic scalene representation, the maximally entangled state was achieved at 62.8 $\mu$s with the decoupled NV center pair $v_{CB} = 5\ kHz$, but for the configuration with the decoupled NV center pair $v_{CB} = 20\ kHz$, the system closely resembled the entangled $\frac{1}{\sqrt{2}}(|000\rangle + |111\rangle)$ state. The fidelity of this state was 0.829, which is not high enough to be considered maximally entangled, but it does point to the potential for disordered systems to reach entangled states at shorter free evolution times. This was seen in the realistic case, where the decoupled pair had a coupling strength of $v_{CB} = 24.1\ kHz$. The short $2\tau_{ent}$ (11.7 $\mu$s) arises from the disorder of the system resulting in a higher frequency of random alignment of the three qubits. Additionally, the coupled pair with $v_{CB} = 24.1\ kHz$ was decoupled during the final period of free evolution, so the stronger coupling $v_{AC(AB)} = 53\ kHz$ dominates the interaction for short $2\tau$, and the weak coupling $v_{AB(AC)} = 4.6\ kHz$ acts to decrease decoherence of the system for short $2\tau$. These factors can allow realistically disordered configurations of coupled NV centers to reach an entangled state at short $2\tau$.

The effective simulation of DEER experiments on coupled NV centers along with previous work simulating hybrid quantum systems and entanglement on a quantum computer [30], demonstrates how simulations performed on IBM Quantum Experience can be used to explore the potential for real NV center spin qubits that can be used in a convenient, accurate quantum computer. Importantly, the free evolution of coupled NV centers can be used to transform multiple NV center spin qubits to desired states. In this case, the desired state was the entangled Bell state, which was achieved at $2\tau_{ent} = 21.2\ \mu s$ for two NV centers with dipolar coupling of $v = 4.93\ kHz$. Since the physical configurations of three coupled NV centers influences the disorder of the system the shortest $2\tau_{ent}$ for three coupled NV centers was different for each of three different configurations: 12.5 $\mu$s for equilateral representation, 6.3 $\mu$s for isosceles representation and 11.7 $\mu$s for scalene representation. Shorter $2\tau_{ent}$ were observed for more ordered systems with a higher coupling constant $R$. The simulations in this work suggest that the entanglement between three strongly coupled NV centers can be achieved for varying symmetric and asymmetric configurations in under 20 $\mu$s, which is much shorter than the coherence time of the individual NV center qubits.

## Discussion

Table 1: Free evolution time required for entanglement for different configurations of 3 coupled NV centers.

| Represented Configuration | | Coupling (kHz) | | | $2\tau_{ent}$ ($\mu$s) | Fidelity |
|---|---|---|---|---|---|---|
| | | $v_{AC}$ | $v_{AB}$ | $v_{CB}$ | | |
| Equilateral | 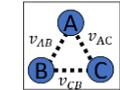 | 50 | 50 | 50 | 12.5 | 0.996 |
| | | 20 | 20 | 20 | 31.5 | 0.996 |
| | | 5 | 5 | 5 | 125.7 | 0.996 |
| Isosceles (Double dominant) | 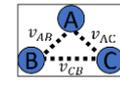 | 50 | 50 | 5 | 62.8 | 0.996 |
| | | 50 | 5 | 50 | 12.5 | 0.896 |
| | | 5 | 50 | 50 | 12.5 | 0.898 |
| Isosceles (Single dominant) | 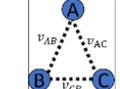 | 5 | 5 | 50 | 6.3 | 0.963 |
| | | 5 | 50 | 5 | 125.7 | 0.990 |
| | | 50 | 5 | 5 | 125.7 | 0.990 |
| Scalene (realistic) | 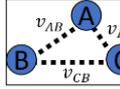 | 53.0 | 4.6 | 24.1 | 11.7 | 0.897 |
| | | 4.6 | 24.1 | 53.0 | 130.1 | 0.972 |
| | | 24.1 | 53.0 | 4.6 | 130.1 | 0.980 |

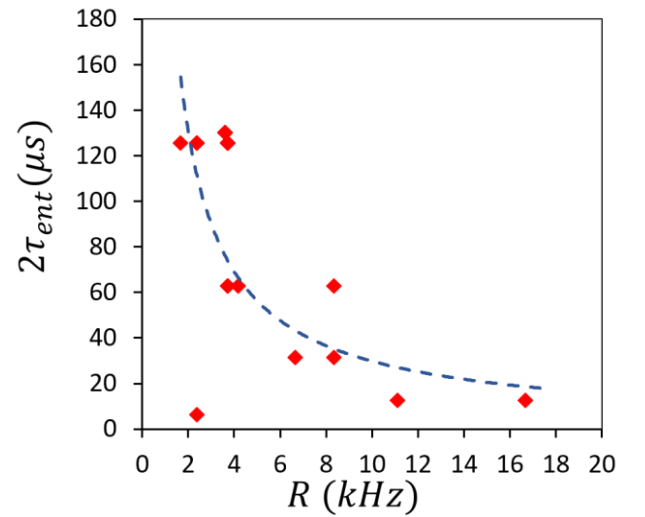

Figure 6: Dependence between $2\tau_{ent}$ and the coupling constant calculated using all three coupled pairs $v_{AC}, v_{AB}, v_{CB}$ as in Eqn. 1.

The coupling constant $R$ (1) provides a way to quantify the order of the system comprised of three coupled NV centers, as well as the order in any pair of coupled NV centers. By neglecting data points with fidelity below 0.9, a trend can be seen in the relationship between the free evolution times required for entanglement and the coupling constant of the system. Figure 3.a clearly shows how the $2\tau_{ent}$ decays as the order ($R$) of the system increases. This trend is most evident when considering all three coupled NV centers instead the case one of the coupled pairs is not considered. This implies that $2\tau_{ent}$ is affected more by the order of the system three NV centers than by the order of each coupled pair. High fidelity entanglement at short free evolution times is therefore favoured by more ordered systems. However, there are some disordered configurations simulated that demonstrate short $2\tau_{ent}$ when simulated, as shown further on in this work.

NV centers created by ion implantation as shown in ref. 18 can form a large number of different coupled configurations and an extended lattice structure with distortions which can be simulated by the present technique. The present work is important from fundamentals of many body interactions between three quantum dots arranged in a form represented by triangles which can be used as the building block for Kondo lattice for observing quantum phase transition when symmetry is broken [2-6]. In general interaction with photons produces sub-radiance and super radiance states where the resonance peak can be tuned with the symmetry of a three-dot configuration. A strong resonance peak relative to non-resonance can be demonstrated by introducing inequalities in interdot couplings through a distribution of the size and distance [23,24]. In some irregular configurations a strong resonant tunnelling can be found from the strong entanglement or a right combination of states as observed by breaking the symmetry of a regular configuration. These special structures would be useful to develop NV center based hybrid quantum devices [31] also by adding squeezed states obtained from strong coupling between a resonator and a couple of NV centers [32]. These closed loop configurations of NV centers can be useful for the development of spin qubits and topological qubits. In summary, the unexpected short $2\tau_{ent}$ for the scalene configuration shows the potential of NV center quantum registers for room temperature operation of quantum computers, because the fabrication of disordered configurations of coupled NV centers is far easier to achieve than the fabrication of ordered systems. Therefore, this is an important step in the development of accessible, convenient and efficient quantum computers that are still accurate.

## Methodology

A double electron-electron resonance (DEER) experiment was simulated to demonstrate the reliability of the $U_c = e^{i\,2\tau\,v_{dip}(S_{zi} \otimes S_{zj})}$ gate. In a DEER experiment, the qubits are both initialised in the $|0\rangle$ state. Applying a $\pi/2$ pulse transforms the sensor qubit into exactly equal superposition: $\frac{1}{\sqrt{2}}(|0\rangle + |1\rangle)$. The qubits are left to evolve freely under the dipolar coupling of the electron spins for a time $2\tau$. A $\pi$ pulse is then applied to the sensor qubit, before being left to evolve freely again for a time $\tau$, after which a $\pi$ pulse is applied to the emitter qubit. After another free evolution period of $\tau$, a $\pi/2$ pulse is applied to the sensor qubit before measurement. The qubits are allowed to interact for increasing free evolution times $2\tau$ [17]. To simulate a DEER experiment, $\pi/2$ microwave pulses are simulated using Hadamard gates and $\pi$ microwave pulses are simulated by NOT gates [17]. The system is initialised in the $|0\rangle$ state for all NV centers, with a measurement gate as the readout operator. The quantum circuit simulating the microwave pulse sequence between time periods of free evolution is shown below. By applying the second $\pi/2$ pulse on both qubits directly after the $\pi$ pulse instead of just before measurement (Figure 6.a), the two qubits can reach the entangled Bell state $\frac{1}{\sqrt{2}}(|00\rangle + |11\rangle)$ after a free evolution time of the order of tens of microseconds. This two-qubit quantum circuit can be extended to entangle three coupled NV centers (Figure 6.b)

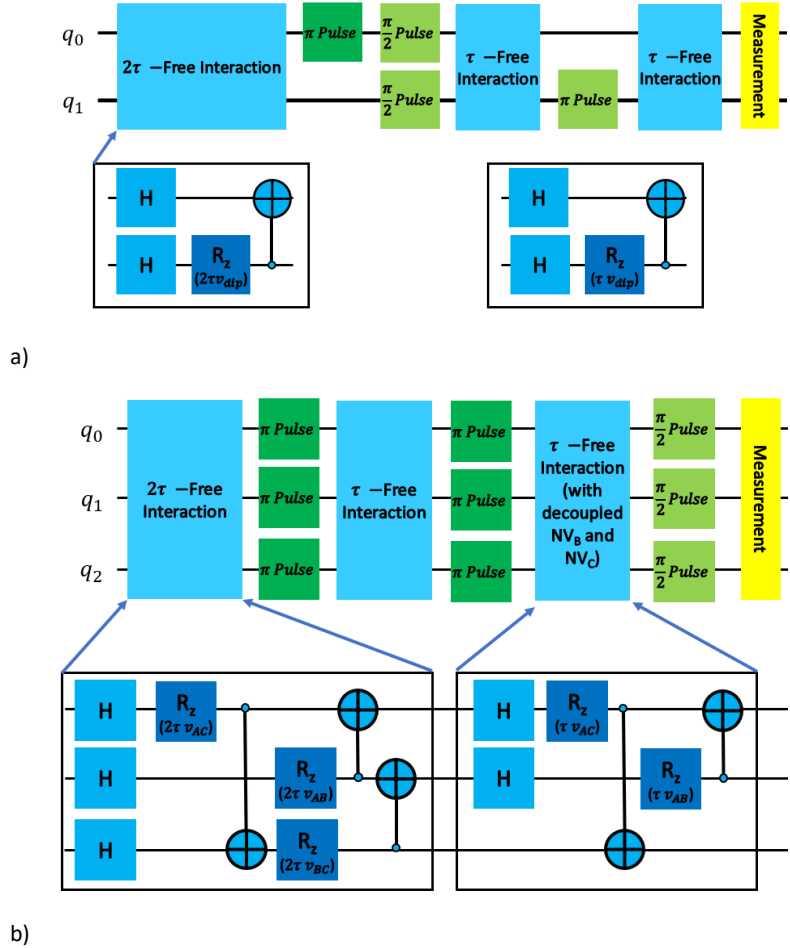

Figure 7: Quantum circuits for entanglement. a) Double electron-electron resonance entanglement pulse sequence b) Triple electron resonance pulse sequence showing the coupling between all three NV centers and the quantum circuit simulating this triple electron resonance scheme. $\pi$ pulses are simulated with a Not gate, $\pi/2$ pulses are simulated with a Hadamard gate.

**Acknowledgements**

We acknowledge use of IBM Q for this work. The views expressed are those of the authors and do not reflect the official policy or position of IBM or the IBM Q team. SB would like to thank D. Churochkin for discussions and NRF (RSA) for financial support.

**References**

[1] Lamm, H., Lawrence, S., Yamauchi, Y., Collaboration, N. *et al.* Parton physics on a quantum computer. *Phys. Rev. Res.* **2**, 013272 (2020).
[2] Barthel, T. Entanglement decomposition for the simulation of quantum many-body dynamics. *Bull. Am. Phys. Soc.* **65,** (2020).
[3] Sharir, O., Levine, Y., Wies, N., Carleo, G. & Shashua, A. Deep autoregressive models for the efficient variational simulation of many-body quantum systems. *Phys. Rev. Lett.* **124**, 020503 (2020).
[4] Kokail, C. *et al.* Self-verifying variational quantum simulation of lattice models. *Nature* **569**, 355–360 (2019).
[5] O'Brien, T. Improvements in quantum algorithms for quantum chemistry and condensed matter. *Bull. Am. Phys. Soc.* **65,** 967 (2020).
[6] Kokail, C., Maier, C., van Bijnen, R., Brydges, T., Joshi, M.K., Jurcevic, P., Muschik, C.A., Silvi, P., Blatt, R., Roos, C.F. and Zoller, P. Self-verifying variational quantum simulation of lattice models. *Nature* **569** (7756), 355-360 (2019).
[7] Bauer, C. W., de Jong, W. A., Nachman, B. & Provasoli, D. A quantum algorithm for high energy physics simulations. *Preprint at https://arxiv.org/pdf/1904.03196.pdf* (2019).
[8] Kuno, Y., Ichinose, I. & Takahashi, Y. Generalized lattice Wilson–Dirac fermions in (1+1) dimensions for atomic quantum simulation and topological phases. *Sci. Rep.* **8**, 1–13 (2018).
[9] Arǵuello-Luengo, J., Gońzalez-Tudela, A., Shi, T., Zoller, P. & Cirac, J. I. Analogue quantum chemistry simulation. *Nature* **574**, 215–218 (2019).
[10] Santagati, R. *et al*. Witnessing eigenstates for quantum simulation of hamiltonian spectra. *Sci. Adv.* **4**, 9646 (2018).
[11] Kawakami, E. *et al*. Gate fidelity and coherence of an electron spin in an si/sige quantum dot with micromagnet. *Proc. Natl. Acad. Sci. U.S.A.* **113**, 11738–11743 (2016).
[12] Nguyen, L. B. *et al*. High-coherence fluxonium qubit. *Phys. Rev. X* **9**, 041041 (2019).
[13] Seo, H. *et al.* Quantum decoherence dynamics of divacancy spins in silicon carbide. *Nat. Commun.* **7**, 1–9 (2016).
[14] Sukachev, D. D. *et al.* Silicon-vacancy spin qubit in diamond: a quantum memory exceeding 10 ms with single-shot state readout. *Phys. Rev. Lett.* **119**, 223602 (2017).
[15] Stoneham, M. Trend: Is a room-temperature, solid-state quantum computer mere fantasy?. *Phys.* **2**, 34 (2009).
[16] Herbschleb, E. *et al.* Ultra-long coherence times amongst room-temperature solid-state spins. *Nat. Commun*. **10**, 1–6 (2019).
[17] Dolde, F. *et al*. Room-temperature entanglement between single defect spins in diamond. *Nat. Phys.* **9**, 139–143 (2013).
[18] Haruyama, M. *et al*. Triple nitrogen-vacancy centre fabrication by $C_5N_4H_n$ ion implantation. *Nat. Commun*. **10**, 1–9 (2019).
[19] Horodecki, R., Horodecki, P., Horodecki, M. & Horodecki, K. Quantum entanglement. *Rev. Mod. Phys.* **81**, 865 (2009).
[20] Wang, Y., Li, Y., Yin, Z.-q. & Zeng, B. 16-qubit IBM universal quantum computer can be fully entangled. *Npj Quantum Inf.* **4**, 1–6 (2018).
[21] Kay, L. E. The evolution of solution state nmr pulse sequences through the eyes of triple-resonance spectroscopy. *J. Magn. Reson.* **306**, 48–54 (2019).
[22] Pribitzer, S., Sajid, M., Hülsmann, M., Godt, A. & Jeschke, G. Pulsed triple electron resonance (trier) for dipolar correlation spectroscopy. *J. Magn. Reson.* **282**, 119–128 (2017).
[23] Bhattacharyya, S. & Churochkin, D. Understanding resonant tunnel transport in non-identical and non-aligned clusters as applied to disordered carbon systems. *J. Appl. Phys.* **116**, 154305 (2014).
[24] Churochkin, D., McIntosh, R. & Bhattacharyya, S. Tuning resonant transmission through geometrical configurations of impurity clusters. *J. Appl. Phys.* **113**, 044305 (2013).
[25] Wu, Y., Wang, Y., Qin, X., Rong, X. & Du, J. A programmable two-qubit solid-state quantum processor under ambient conditions. *Npj Quantum Inf.* **5**, 1–5 (2019).
[26] Hirose, M. & Cappellaro, P. Coherent feedback control of a single qubit in diamond. *Nature* **532**, 77–80 (2016).
[27] Shkolnikov, V., Mauch, R. & Burkard, G. All-microwave holonomic control of an electron-nuclear two-qubit register in diamond. *Phys. Rev. B* **101**, 155306 (2020).
[28] Kim, M. *et al*. Decoherence of near-surface nitrogen-vacancy centers due to electric field noise. *Phys. Rev. Lett.* **115**, 087602 (2015).
[29] IBM Quantum Experience. https://quantum-computing.ibm.com/ (2020).
[30] Ajoy, A., Bissbort, U., Poletti, D. & Cappellaro, P. Selective decoupling and Hamiltonian engineering in dipolar spin networks. *Phys. Rev. Lett.* **122**, 013205 (2019).
[31] Mazhandu, F., Mathieson, K., Coleman, C. & Bhattacharyya, S. Experimental simulation of hybrid quantum systems and entanglement on a quantum computer. *Appl. Phys. Lett.* **115**, 233501 (2019).
[32] Mathieson, K. and Bhattacharyya, S. Hybrid spin-superconducting quantum circuit mediated by deterministically prepared entangled photonic states. *AIP Adv.* **9**, 115111 (2019).